\documentclass[10pt]{article}
\usepackage{journal_abbr_mac,graphicx,natbib}

\title{``Grandeur in this view of life": N-body simulation models of the Galactic habitable zone}
\author{B. Vukoti\'c$^{1}$\thanks{E-mail:
bvukotic@aob.rs (BV), Dominik.Steinhauser@uibk.ac.at (DS), gerardo.martinez.aviles@gmail.com (GMA), mcirkovic@aob.rs (MMC), micic@aob.rs (MM), sabine.schindler@uibk.ac.at (SS).}, D. Steinhauser$^{2}$, G. Martinez-Aviles$^{3}$, M.M. \'Cirkovi\'c$^{1}$,\\ M.Micic$^{1}$ and S. Schindler$^{2}$}
\begin{document}

\maketitle

{\small
$^{1}$Astronomical Observatory, Volgina 7, 11060 Belgrade 38, Serbia\\
$^{2}$ Institute of Astro- and Particle Physics, University of Innsbruck, Technikerstr. 25, 6020, Innsbruck, Austria\\
$^{3}$ Observatoire de la C\^ote d'Azur - Boulevard de l'Observatoire - CS 34229 - F 06304 NICE Cedex 4, France
}


\begin{abstract}
We present an isolated Milky Way-like  simulation in GADGET2 N-body
SPH code. The Galactic disk star formation
rate (SFR) surface densities and stellar mass indicative of Solar neighbourhood are
used as thresholds to model the distribution of stellar mass in
life friendly environments. SFR and stellar component density are
calculated averaging  the GADGET2 particle properties on a 2D
grid mapped on the Galactic plane. The peak values for possibly
habitable stellar mass surface density move from $10$ to $15$ kpc
cylindrical galactocentric distance in $10$ Gyr
simulated time span. At $10$ Gyr the simulation results imply the following. Stellar particles which have spent almost all
of their life time in habitable friendly conditions reside typically at
$\sim16$ kpc from Galactic centre and are  $\sim 3$ Gyr old.  Stellar particles that have spent $\ge 90 \%$ of their $4-5$ Gyr long life time in habitable friendly conditions, are also predominantly found in the outskirts of the Galactic disk. Less then $1 \%$ of these particles can be found at a typical Solar system galactocentric distance of $8-10$ kpc. Our results imply that the evolution of an isolated spiral galaxy is likely to result in galactic civilizations emerging at the outskirts of the galactic disk around stellar hosts younger than the Sun.
\end{abstract}

\section{Introduction}
Boosted by the recent discoveries of the ever increasing number
of extrasolar planets and their candidates \citep{howard13},
Galactic habitability studies are gaining momentum and are becoming an
important part of the mainstream astrobiological research
\citep{2014MNRAS.440.2588Spitoni_etal,Dayal15,2015arXiv151101786Forgan}.
Apart from Earth, biospheres still remain unobserved, both inside and
 outside the Solar System. The absence of the empirical data causes an extreme
difficulty for the robust studies on the evolution of living matter in the Universe. This is not likely to change until the arrival of the capacity to detect biomarkers from interstellar distances in various extrasolar planetary systems across
the Galaxy. Since the matter in the Universe is primarily organized into galaxies, the
evolution of life in the Universe will strongly correlate with
the evolution of galaxies. 

The present computing capabilities to simulate Galactic evolution
can be very useful for understanding the significance of
different Galactic processes in shaping the spatio-temporal
distributions of Earth-like habitats. Following the introduction
of the Galactic Habitable Zone (GHZ) concept
\citep{Gonzalez_et_al_2001Icar} we have further developed a
general idea what are the basic preconditions for
terrestrial-like habitats to appear and remain habitable. Apart
from the chemical evolution, galactic dynamics can be of great
importance for habitability in spiral galaxies due to radial
population mixing \citep{Roskar_et_al_2011ASPC, 2008ApJ...684L..79R}. This can
significantly influence the distribution of habitable systems in
the Galactic disk. In the previous numerous models of the Galactic
habitability and SETI-oriented studies
\citep[i.e.,][]{Lineweaver_et_al_2004Sci,
Gowanlock_et_al_2011AsBio, Forgan_2009IJAsB8121F,
Cotta_Morales2009JBIS, Prantzos2013IJAsB,
2012OLEB...42..347V}, stellar dynamics was not accounted for
until the recent work of \citet{2015arXiv151101786Forgan}. The
main goal of this paper is to determine the allowed
habitable times forced upon the stellar systems from star
formation rate (SFR) and stellar density as the global habitability
agents in the Galaxy. We present and analyse an N-body SPH
simulation of a single isolated Milky Way-like galaxy in order to model the spatio-temporal
distribution of stellar mass in life friendly environment. 

This has even wider scientific and methodological importance. In the famous last paragraph of "The Origin of Species"
\citet{Darwin59} contrasted apparent simplicity of Newton's law of
gravity and the consequent Keplerian motion of planets with the
ecological complexity of "an entangled bank" created by
biological evolution. In the very last sentence, however, he
affirmed the underlying idea that {\it both\/} these aspects of
nature, astronomical and biological, are explicable in the
naturalistic terms by law-like dynamical regularities -- and that
we find "grandeur in this view of life" evolving from simple to
multiple complex forms. The quantitative sort of astrobiological
work as presented here is, in more than one sense, honouring and
continuation of this Darwinian programme.

In the rest of this introductory section we describe the basic
habitability factors that are considered in analysis of our
simulation. Simulation is described in Section 2, details of
the model are given in Section 3, while discussion and summary of
the results are given in sections 4 and 5, respectively.

\subsection{Habitability agents}
Explosions of nearby supernovae are likely to be hazardous for a
life dwelling planet up to a few pc distance
\citep{Gehrels_etal2003ApJ} or even up to a kpc \citep{Karam02}.
At Galactic distance scales they can be considered as a local
phenomena and their life hazard potential treated accordingly.
An overall danger from a supernova is likely to be
smaller compared to a stellar collision induced hazards and an
occasional stress from a nearby supernovae can even boost the
biological evolution \citep{Filipovic_etal2013SAJ}. However, a planetary
system in the midst of a star forming region characterized by a
high supernovae rate is likely to be hindered from developing a
"fruitful" biosphere.

{\it Continuity of habitable conditions\/} is a necessity
prerequisite for a planet to develop a versatile biosphere.
Although minor dynamically stable changes can be suspended with
plate tectonics feed-back cycles, any sufficiently large
perturbation can push the planet outside the habitable
temperature range without the possibility to recover. (It has
almost happened on Earth as well during at least two "Snowball
Earth" episodes during the Precambrian supereon.) A habitable
planet should thus be as close as possible to a circular orbit
around its host star. For multiple stellar systems,
configurations with widely separated stellar components (where a
planet is mostly bound and in proximity of a single stellar
object) or very close in (where a planet revolves around multiple
stellar host) are preferred. In crowded regions of the Galactic
disk a frequent nearby passages (at a distances of $\sim 100$ AU)
of neighbouring stars may cause a large stress on the orbits of
their planets, either by disruption of planetary orbits  or by
perturbation of cometary and asteroidal belts. This can cause a life-devastating flux of impactors. In close enough encounters a large
perturbation can even permanently push the planet outside
circumstellar habitable zone or set it free from the host stellar
system gravitational grasp.

Changes in the
atmosphere of a life bearing planet, caused by a nearby stellar explosion or a stellar
collision caused orbital disruption, might turn a planet into a "snowball", or
a Venus-like world. Of course, this does not mean that the planet will become
completely lifeless. Extremophiles, such as a single cell
organisms that live in extreme conditions here on Earth, can
still be present. However, the potential for evolution of life on
such a planet would probably be severely constrained due to a very
small number of ecological niches available
\citep{conwayMorris2003}. Such a fair biosphere is less likely to
produce a complex intelligent species, detectable  via
interstellar travel, distant communication and other means, such
as IR excess from dissipated energy
\citep{2015ApJ...810...23Zackrisson}. There are multiple reasons
to believe that evolution of cognition and intelligence is
possible only within a large and diverse habitats, where many
necessary evolutionary innovations could be achieved
\citep{1995ASPC...74..143Russell,Vermeij07022006_2006,Morris555_2011}.
It is unlikely to happen in tightly constrained,
ultra-specialized habitats.

It is quite obvious that the existence of rocky planets is
conditional upon presence of their building blocks in the
interstellar matter, metals. Apart from heavy nuclei that sink
down to cores of planetary objects during the cooling stage of
their formation, medium weight metals such as silica, magnesium,
carbon, nitrogen, oxygen, etc, are main constituents of mantle
and crust. At first look it seems that higher metallicity of
interstellar matter  will yield higher number of rocky planets,
but this need not be the case. The "goldilocks" approach taken by
\cite{Lineweaver2001Icar, Pena_CabreraDurand_Manterola04} relies
on the assumption that high metallicity is likely to lead to
formation of gas giants with rocky cores, preferentially via core
accretion model \citep[see also,][]{2012Natur.486..375Buchhave, 2000ApJ...537.1013I}. The samples of
exoplanets show that the highest number of discovered systems
have hosts with near Solar metallicity. While this strongly
favours habitability in the "goldilocks" manner, the existence of
correlation between metallicity and the number of discovered
rocky planets is far from clear since there are strong selection effects in play. Recent theoretical results
indicate that planet formation starts at lower values of
metallicity than previously assumed, already at $Z \geq 0.1 \,
Z_\odot$ \citep{2012ApJ...751...81Johnson_Li}. The small number
of discovered Earth-like planets and higher probability of
detecting gas giants with current observational techniques are
main reasons to question the existence of such a correlation
\citep[see][and references therein]{Prantzos2008SSRv}. For the
purpose of our model we constrain the metallicity of the possible
stellar particles considered for habitability to be above the
lowest determined values for existing samples of Earth-like
exoplanets.

\section{Simulation details}
\label{sim}

Following the model presented in
\citet{Springel_Hernquist2003MNRAS}, a small-scale simulation of
an individual star-forming disk galaxy in GADGET2 code
\citep{2001NewA....6...79Springel_Yoshida_White,
2005Natur.435..629Springel_etal} was performed. The halo was set
up in isolation following the approach taken by
\citet{Navarro_etal1997ApJ}, with the gas and dark matter
initially in virial equilibrium, known as the Navarro, Frenk \&
White (NFW) halo:
\begin{equation}
\frac{\rho(r)}{\rho_\mathrm{c}}=\frac{\delta_\mathrm{c}}{(r/r_\mathrm{s})(1+r/r_\mathrm{s})^2},
\end{equation}
where $r_\mathrm{s}$ is the halo scale radius, $\delta_\mathrm{c}$ is a characteristic (dimensionless) density, and $\rho_\mathrm{c} = 3H^2/8\pi G$ is the cosmological critical density.
The conventional parameter in this kind
of model \citep{Binney_Tremaine2008gady} is known as $r_\mathrm{200}$, which is the distance from
the centre of the halo at which the mean density is 200 times the $\rho_\mathrm{c}$. The mass interior $M_\mathrm{200} = 200 \frac{4}{3} \pi r^3 \rho_\mathrm{c}$ was chosen to be $M_\mathrm{200} =10^{12} M_\odot$, being baryonic 10\% of the mass. For the halo concentration factor $c=r_\mathrm{200}/r_\mathrm{s}$ a value of $c=9.0$ was chosen. To describe the initial angular momentum $J$ of the halo, the spin parameter $\lambda = \frac{J|E|^{1/2}}{GM_\mathrm{vir}^{5/2}}$ is usually used. To produce a large disk, a value of the spin parameter $\lambda = 0.1$ was chosen.

For the purpose of  simulating radial migrations in the galactic disk \citet{2008ApJ...684L..79R} used gas particles of $10^5 \mathrm{M}_\odot$ and stellar particles of $3\times10^4 \mathrm{M}_\odot$. In the initial conditions, we use $10^6$ gas particles with a mass resolution of $10^5 \mathrm{M}_\odot$. The gravitational softening length was set to $0.8$ kpc for both, gas and stellar particles. We include radiative cooling \citep{1996ApJS..105...19Katz_etal} and recipes for star formation and stellar feedback as described in \citet{Springel_Hernquist2003MNRAS}. Each gas particle can produce up to two stellar particles of $5\times10^4 \mathrm{M}_\odot$; When a gas particle produces the second stellar particle it is used up and is not in the simulation any more. A fixed metallicity value is assigned to the stellar particles equal to the metallicity of their gas particle progenitor at the time when the stellar particle is produced. After a $10$ billion
years simulation, there were $1055078$ stellar particles, and $488158$ gas particles. We sampled a $1000$ snapshots from the simulation which translates into a $10^7$ yr time resolution.

\section{Habitability model}

There are $\sim 10^{11}$ stellar systems in a typical $L_\ast$ galaxy. Even with the developed computing hardware of today, N-body simulations of more than  $\sim 10^{7}$ particles are a very rare occurrence. Usual mass resolution for a stellar component is in the $ 10^{4}-10^{6}\mathrm{M}_\odot$ interval. At present level of available computing resources it is not possible to make a full scale N-body simulation of the Galaxy with mass resolution at the level of the individual stellar systems. This makes it difficult to estimate the habitability of the individual stellar system. The global Galactic parameters such as SFR or stellar density can be modelled and calculated with high fidelity. Still, it  is  not possible to track the individual stellar system during the simulation and examine its position for an environmental habitability. At the level of resolved stellar mass, in a typical N-body simulation of the galaxy, stellar particles usually have masses of $\sim10^5\mathrm{M}_\odot$. This is more representative of a stellar cluster, and that mass should be considered as smeared out rather than concentrated in a given position of the stellar particle.

The habitability of the stellar particles (see section \ref{sim}) is calculated from the relevant Galactic parameters on a spatial 2D grid. The grid is mapped on the Galactic disk (xy plane in the simulation xyz coordinates). The 2D model requires less computation time than a more realistic 3D approach. Also, the results representation is straightforward since the plane of the disk is represented by the plotting plane. A significant precision improvement cannot be achieved with the 3D approach at the current level of the simulation performance, since the softening length in our simulation is of the order of thick disk thickness.

\begin{figure*}
\includegraphics[width=1.0\textwidth]{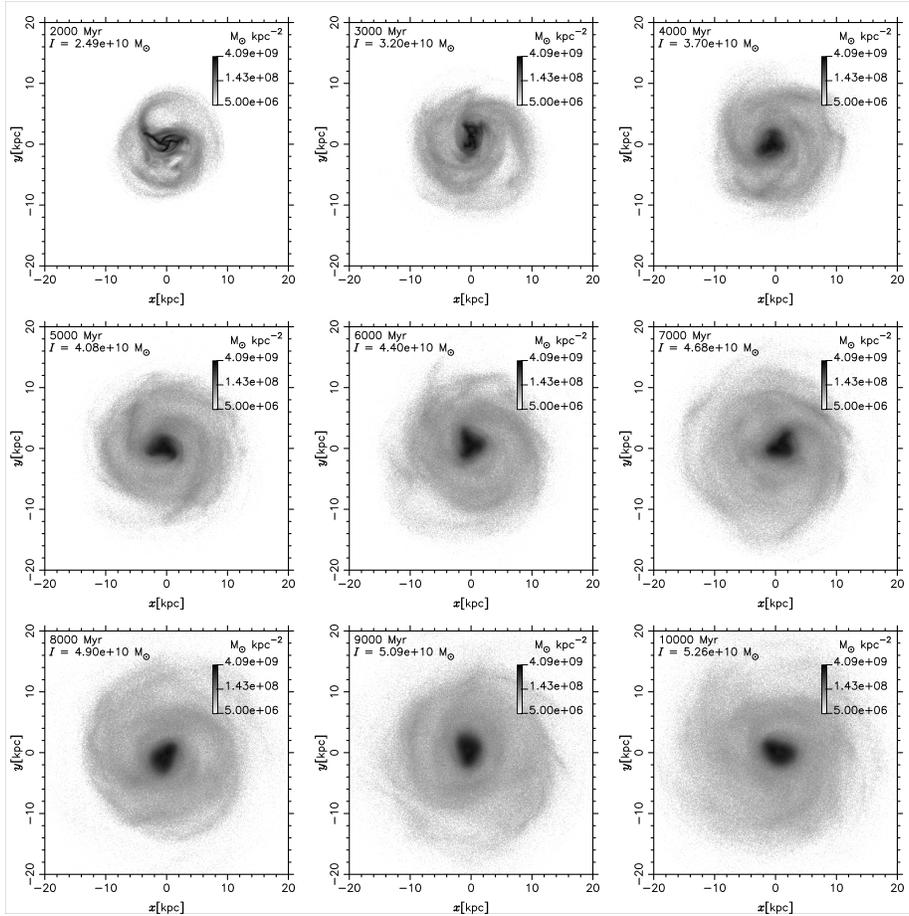}
\caption{Stellar mass surface density. The time instant and integrated value for the given plotting range are indicated in each panel.}
\label{stelar_density}
\end{figure*}

\begin{figure*}
\includegraphics[width=1.0\textwidth]{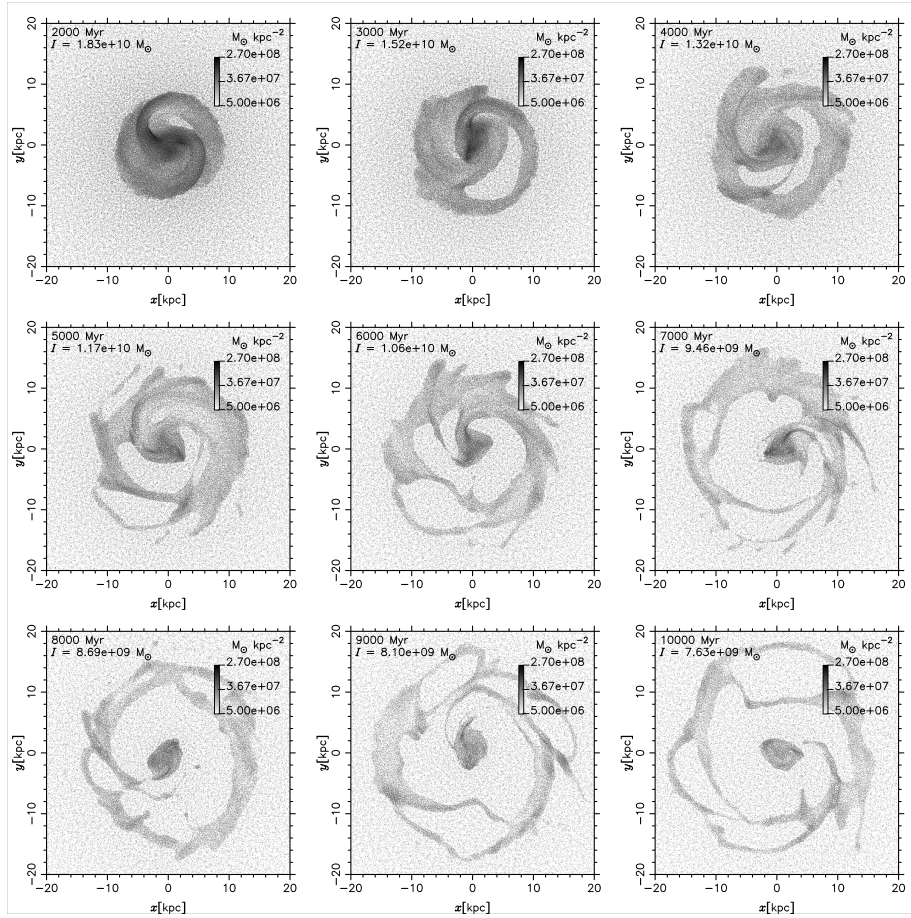}
\caption{Gaseous mass surface density. The time instant and integrated value for the given plotting range are indicated in each panel.}
\label{gas_density}
\end{figure*}

\begin{figure*}
\includegraphics[width=1.0\textwidth]{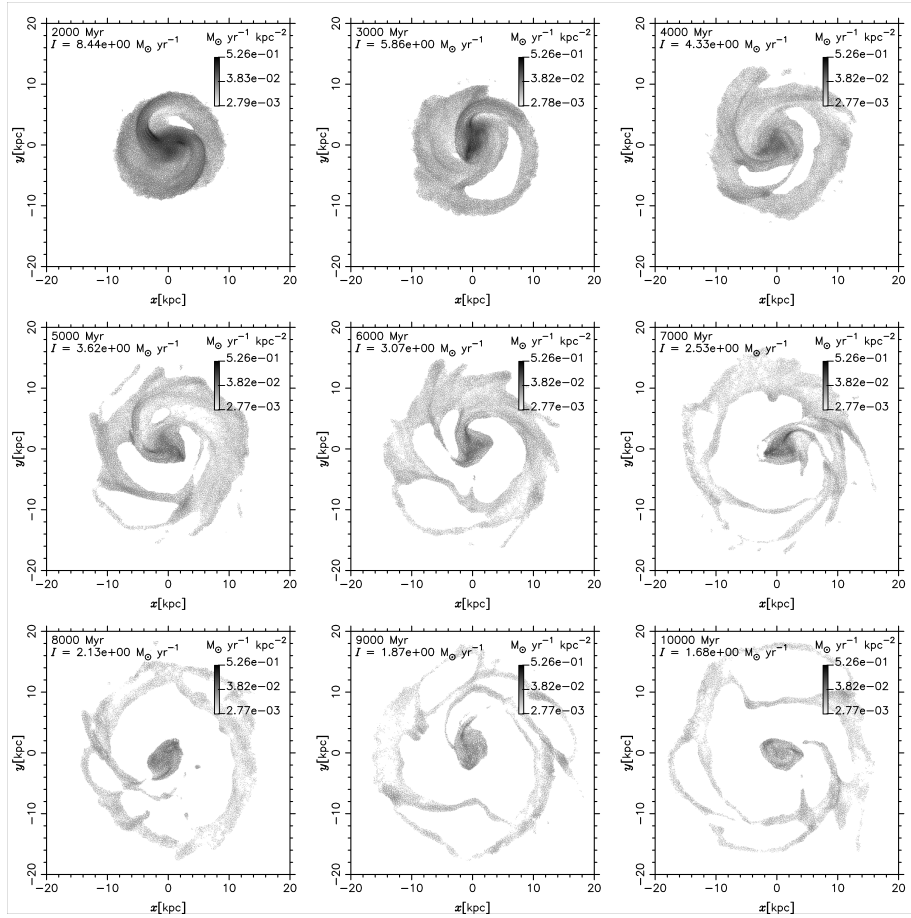}
\caption{Star formation rate. The time instant and integrated value for the given plotting range are indicated in each panel.}
\label{sfr}
\end{figure*}

The SFR for each grid cell is estimated as the sum of the SFRs of the individual gas particles located within the cell. In a similar fashion a density of stellar (and gaseous)  component is estimated from stellar (gas) particles found in a grid cell at the moment of calculation. The resulting surface densities for SFR, stellar and gaseous mass, are given in Figures \ref{sfr}, \ref{stelar_density} and \ref{gas_density} respectively. It is evident that the star formation activity closely matches the dense gas areas. Since the modelled Galaxy is isolated, after about $3$ Gyr SFR significantly drops. At latter times the areas of intense star formation are not solely found in the bulge of the Galaxy and inner disk but are drifting towards the disk outskirts as an expanding arms of the fading spiral pattern. The inner disk is thus left with little to no areas of the intense star formation. Unlike gas, very dense areas of the stellar component are confined to vicinity of the Galactic Centre at all times. The denser areas in the disk arise as a loose spiral pattern that is expanding as more particles from the interior drift away. This gradually expands the stellar disk from $\sim  10$ to $\sim 15$ kpc.


We constructed an analytical  habitability recipes to quantify the habitability relevant properties of the Galactic disk. These recipes are applied on the simulation data. In the remainder of this section this ``post-simulation" analysis of ours is described in detail.

\subsection{Habitable environment}
The habitability relevant parameters are calculated from the characteristics of the individual particles on a $400 \times 400$ grid in the XY simulation coordinate plane. The grid resolution is $100$ pc and $100$ pc along the x and y-axis, respectively.  The grid cells are considered as the habitable friendly environment for the particles within them, if calculated values for the stellar number density are below $6.1\times10^{7}~\mathrm{M}_{\odot}~\mathrm{kpc}^{-2}$ \citep[following][]{2015arXiv151106387McGaugh} and the SFR is lower than $3.0\times10^{-3}~\mathrm{M}_{\odot}~\mathrm{kpc}^{-2}\mathrm{yr}^{-1}$ \citep[consistent with][]{2002NewA_DeDonderVanbeveren}. These values are selected as indicative of the Solar neighbourhood.

\subsection{Metallicity}
\label{habmet}
The basic precondition for the existence of a rocky planet is the sufficient amount of heavy elements available in the proto-planetary disc during planet formation. On the other hand, a reasonable and generally well accepted hypothesis is that a higher amount of metals in the proto-planetary disc leads to a larger probability of having gas giants. The gas giants can extensively perturb and even destroy the smaller, solid surface planets \citep[i.e, see][]{Lineweaver2001Icar}. However, the metallicity at which probability of the gas giants formation is significant for the habitability of the host stellar system cannot be determined with certainty. This is caused by observational selection effects. The giant planets are more likely observed then smaller ones and the host star atmosphere is contaminated with debris from the proto-planetary disc. In the work of \citet{2012Natur.486..375Buchhave} it is implied that the Earth-size exoplanets form around stars with a wide range of metallicities while the giant planets are more common around high metallicity hosts. On the other side of the metallicity scale the absence of building blocks means no existence of the Earth-like habitats at all.

The metallicity is expressed as $[\mathrm{F_e}/\mathrm{H}]$, the $\mathrm{F_e}$ relative to hydrogen abundance, on a logarithmic scale where zero is the Sun's metallicity. We adopted a Gaussian probability distribution for the metallicity dependent habitability criteria. The distribution mean value is at $[\mathrm{F_e}/\mathrm{H}]=-0.075$ with standard deviation ($\sigma$) of $0.1125$. This way, the $2\sigma$ lower end of the selected distribution is at $[\mathrm{F_e}/\mathrm{H}]>-0.3$. This is the lowest metallicity in the sample of 7 well measured stellar systems with the Earth-size exoplanets \citep{2015ApJ...815....5S}. The distribution is normalized so that the highest value of the probability density is $1.0$. A pseudo-random number \citep{SaitoMatsumoto08} from a $(0,1)$ interval, with a uniform density, is assigned to each stellar particle in the simulation. If the assigned number is smaller than the corresponding value of the normalized Gaussian probability density distribution, the particle is considered as a potential habitat. The results are averaged over five separate runs of our habitability analysis.  The more simple case, of using a top-hat probability function instead of the Gaussian, gives similar results. The only significant difference is that the number of the potentially habitable particles is larger. This is expected since in a top-hat case all particles with metallicity higher than $-0.3$ represent the potential habitats.

\subsection{Habitable time}
\label{habtime}
One of the most important parameters for the research of the astrobiological development of a galaxy is an estimate of the time span an individual stellar system spends continuously dwelling in a habitable friendly environment. However, the simulation finite time resolution prevents from calculating habitability relevant Galactic parameters at each point of the particle trajectory. The higher the sampling rate along the particle trajectory ($\mathbf{=}$ smaller time step of the analysis) the higher the chances that the particle can be found in life inhospitable conditions. This means that a higher sampling rate will reduce the number of continuously habitable particles for a given biological time scale. As a consistent temporal measure of particle habitability we adopt:
\begin{equation}
f_\mathrm{ht} = \frac{t_\mathrm{h}}{a},
\end{equation}
where  $t_\mathrm{h}$ is the number of snapshots for which the particle was found in a habitable grid cell, and $a$ is the total number of snapshots that the stellar particle spent in the simulation.  We adopted the following habitability preconditions for the stellar particles:
\begin{itemize}
\item[$\star$)] $t_\mathrm{a}  > 1~\mathrm{Gyr}$ and\\
\item[$\star$ $\star$)] the particle is the potential habitat according to metallicity criteria given in Section \ref{habmet},
\end{itemize}
where $t_\mathrm{a}$ is the age of the particle, calculated as number of snapshots that the particle spent in the simulation multiplied by the temporal resolution (time step of the post simulation analysis, $\tau$).
In the following, we will refer to the stellar particles that satisfy the above conditions as candidate habitable particles (CHPs). The stellar particle is considered habitable (HP) if in addition to conditions ($\star$) and ($\star$ $\star$) it also satisfies:
\begin{equation}
f_\mathrm{ht} > 0.5.
\label{h3}
\end{equation}
The $t_\mathrm{a}>1$ Gyr condition is set so that only the particles with (on average) fully developed and biologically active planets are considered. The Hadean eon on Earth lasted for 0.6 Gyr and the oldest indirect evidence for life on Earth is dated  back to the end of the Hadean and beginning of the Archean eon. Also, 1 Gyr gives enough time so the $f_\mathrm{ht}$ is less susceptible to statistical fluctuations.

\begin{figure*}
\includegraphics[width=0.45\textwidth]{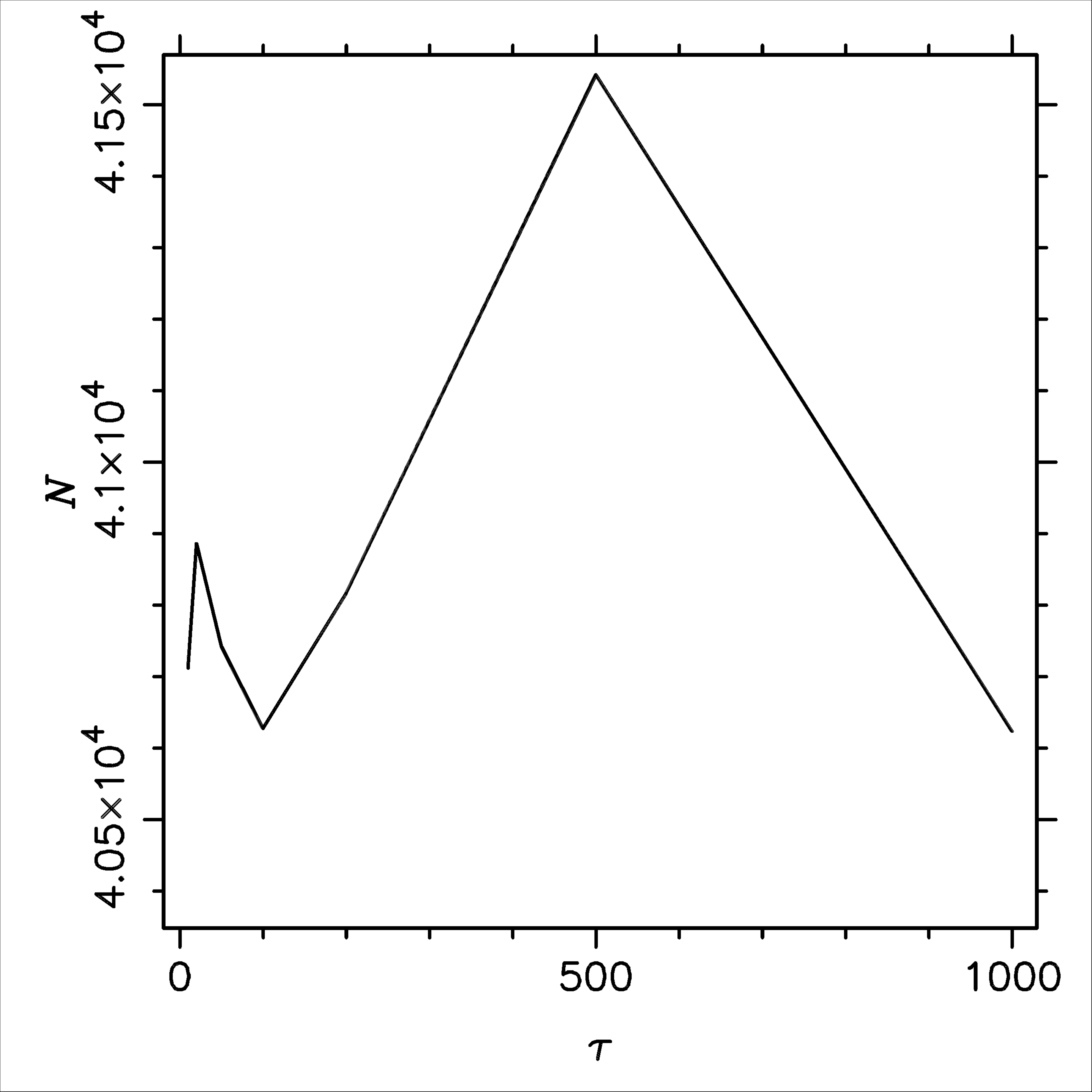}
\includegraphics[width=0.45\textwidth]{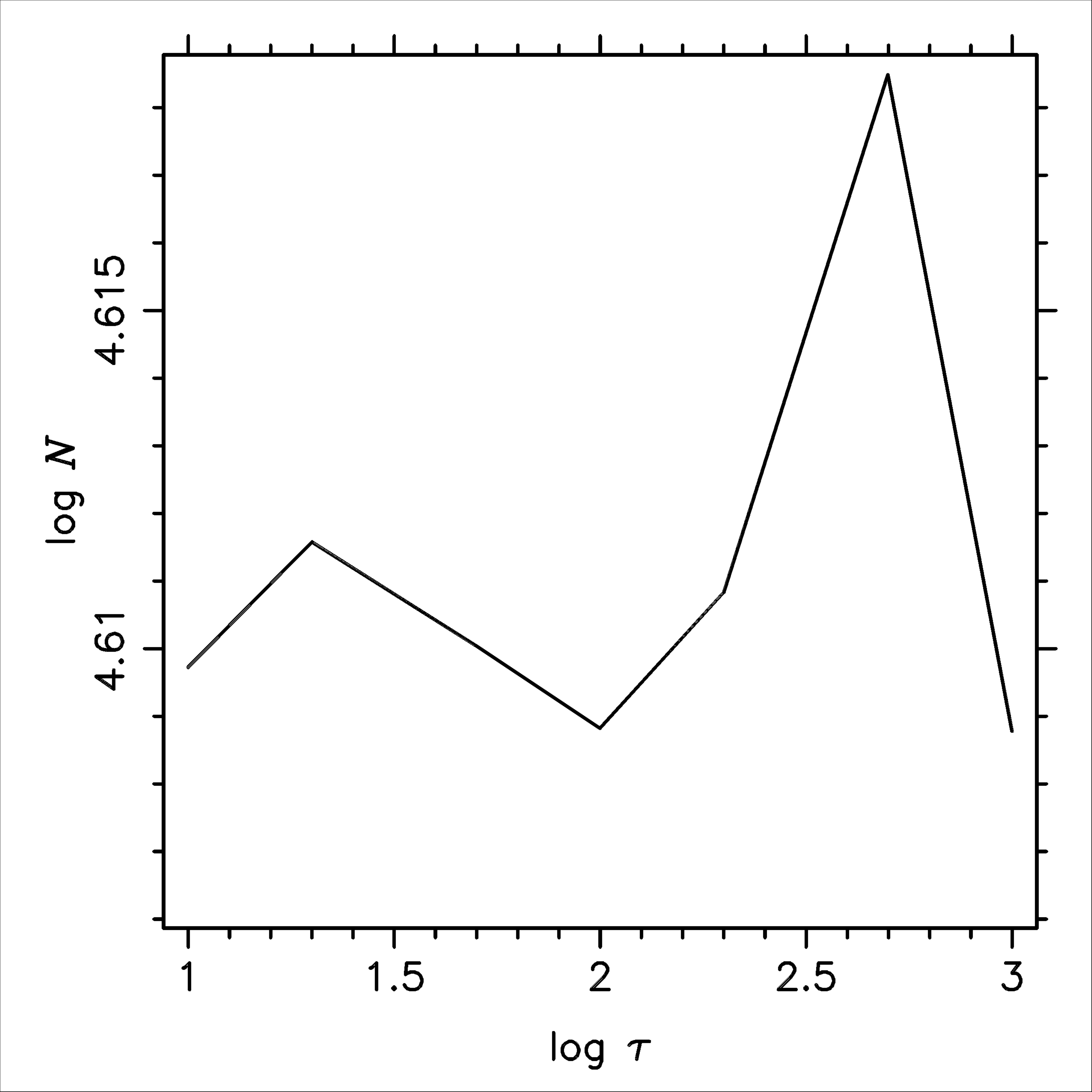}\\
\caption{Number of habitable particles convergence for different time steps at the end of the simulation, $10$ Gyr. Values are averaged over 5 runs.}
\label{convergence}
\end{figure*}

To examine the consistency of our habitability time estimates, we test for the convergence of the HPs number. At $10$ Gyr we plot the number of HPs ($N$), against the variable time step $\tau$ used in the convergence test. The selected values were $\tau=\{1000, ~500, ~200, ~100, ~50, ~20, ~10\} $ Myr. The test results are presented in Figure \ref{convergence}. The slope change on the linear scale plots clearly varies in sign (although apparently less evident on logarithmic plot). We cannot make any high probability conclusions even if smaller values for $\tau$ were to be included in the test (which will definitely have to account for much larger processing times of the simulation data analysis). However, the span of the $N$-axis is $\sim2\%$ of the presented $N$ values (even smaller when only  small $\tau$ values are considered). This adds confidence to the conclusion that the constructed measure of habitability $f_\mathrm{ht}$ is robust and reliable.

\section{Results and discussion}

\begin{figure*}
\includegraphics[width=1.0\textwidth]{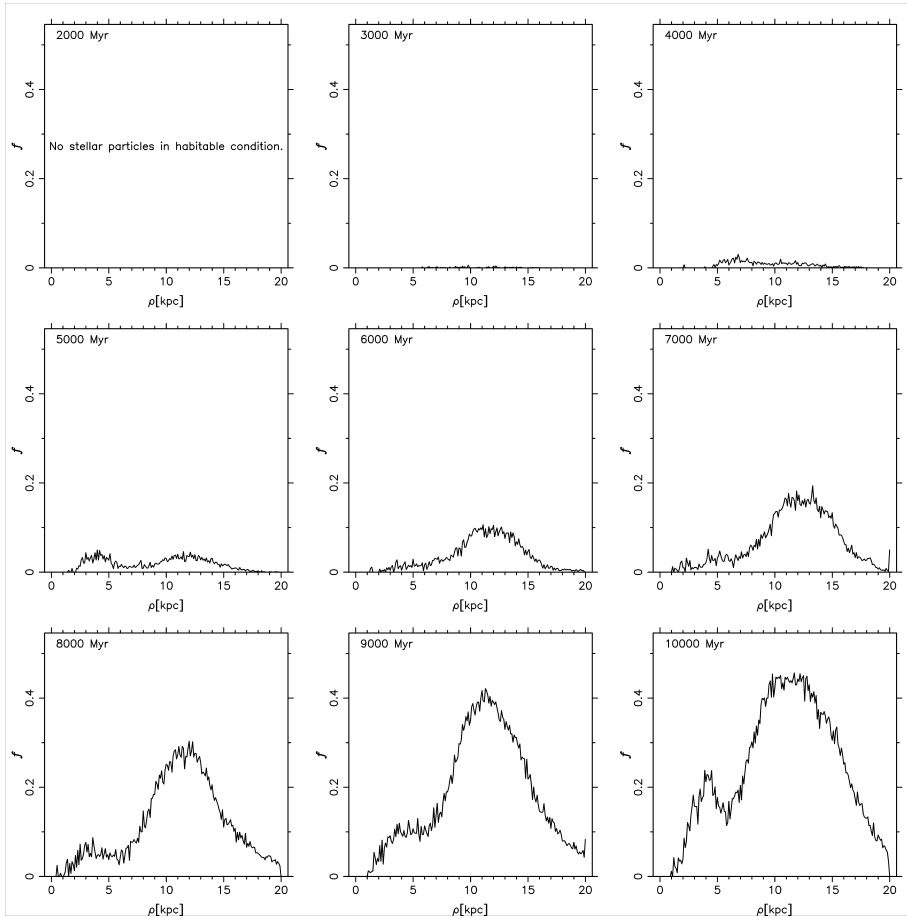}
\caption{Distribution of fraction of the habitable grid cells (containing at least one HP) at a given radius $\rho$. Averaged over 5 runs.}
\label{hc_frac_rho}
\end{figure*}

\begin{figure*}
\includegraphics[width=1.0\textwidth]{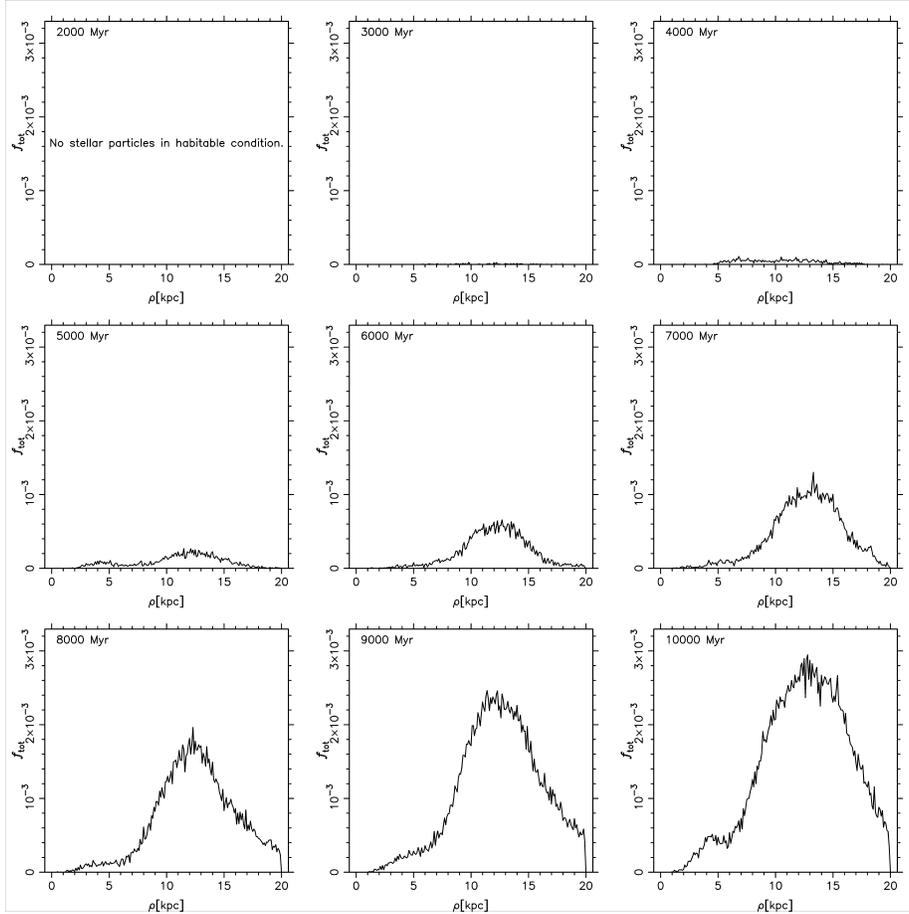}
\caption{$\rho$ distribution for the fraction of the habitable grid cells (containing at least one HP) from total number of grid cells for $\rho < 20$ kpc. Averaged over 5 runs.}
\label{hc_frac_total}
\end{figure*}

\begin{figure*}
\includegraphics[width=1.0\textwidth]{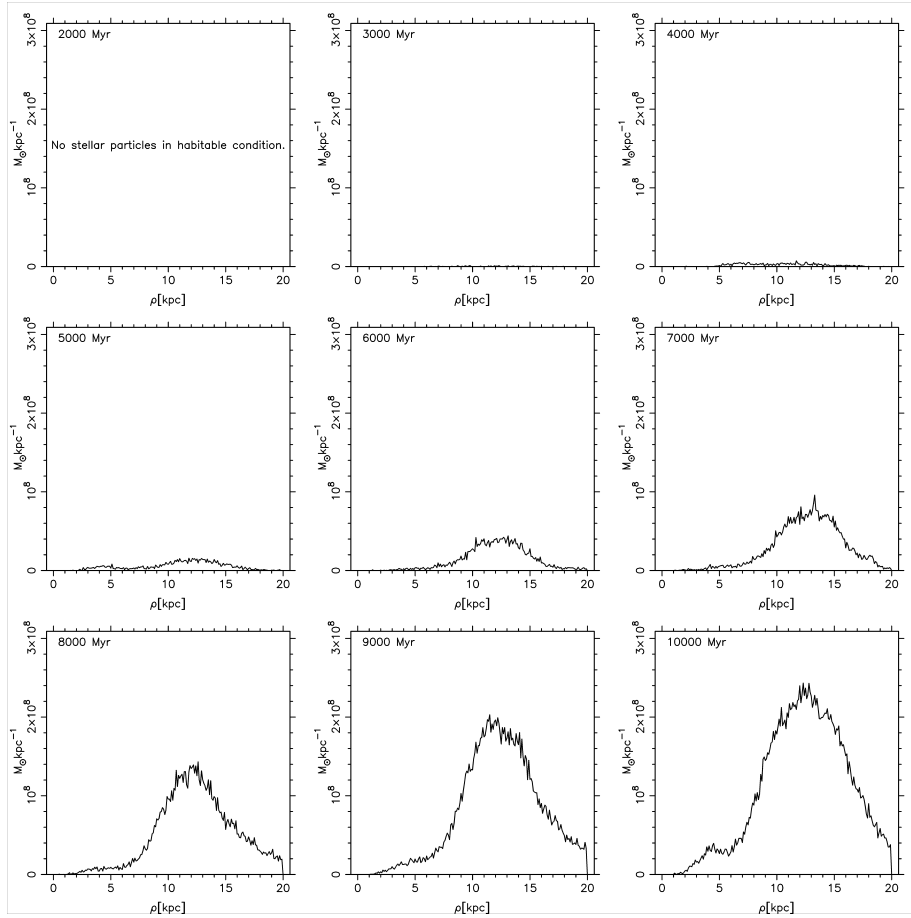}
\caption{Habitable stellar mass (particles) per unit galactocentric distance. Averaged over 5 runs.}
\label{hp_rho}
\end{figure*}

\begin{figure*}
\includegraphics[width=1.0\textwidth]{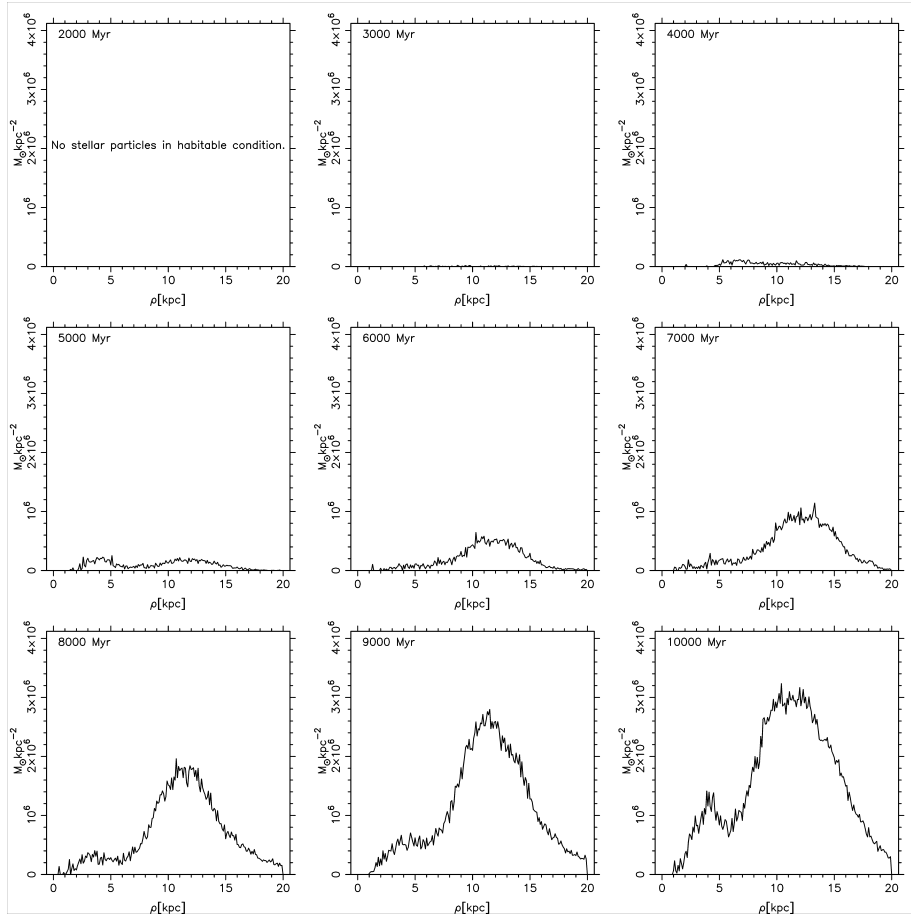}
\caption{Surface density of habitable stellar mass (particles) averaged over angular coordinate in the plane of the disk for a given $\rho$. Averaged over 5 runs.}
\label{hp_2pirho}
\end{figure*}

Inspection of Figure \ref{hc_frac_rho} shows that after the initial $\rho \sim 5$ kpc peak  at $\sim 5$ Gyr, the $10-15$ kpc range has the highest fractional (and absolute, Figure \ref{hc_frac_total}) spatial coverage with HPs at the second half of the simulated time span. From Figures \ref{hp_2pirho} and \ref{hp_rho} it is evident that distribution of HPs follows in shape the distribution of disk areas containing HPs. Following the results of \citet{Roskar_et_al_2011ASPC}, that emphasize outward migrations caused by dynamical friction, we make the following interpretation.  The distribution peak at $5$ kpc at earlier times is probably caused by the build up of stellar particles formed from dense gas areas near the bulge (Figure \ref{sfr}). In the second half of the simulated time span the presence of high SFR areas around $\rho = 10$ kpc likely causes the appearance of a significant number of new stellar particles. These particles are then scattered preferably to the outside of the disk, (Figures \ref{sfr} and \ref{stelar_density}) contributing to the peaks of the habitable stellar mass and the habitable disk surface areas at $\rho = (10,15)$ kpc.

From  Figures \ref{hc_frac_rho} and \ref{hp_2pirho} at $10$ Gyr,  $\approx 3 \times 10^6~\mathrm{M_\odot}\mathrm{kpc}^{-2}$ is distributed over $\approx 40\%$ of the grid cells (for $10-15$ kpc range) and $\approx 1.5 \times 10^6~\mathrm{M_\odot}\mathrm{kpc}^{-2}$  is distributed over $20\%$ of the grid cells (at $\approx 5$ kpc). This gives $\approx  7.5 \times 10^6~\mathrm{M_\odot}\mathrm{kpc}^{-2} $, which translates to $\approx 7.5 ~\mathrm{M_\odot}\mathrm{pc}^{-2}$ as the average surface density of the habitable stellar mass in the habitable disk areas. This value is somewhat higher than the lowest value for the stellar component density on Figure \ref{stelar_density} ($5 ~\mathrm{M_\odot}\mathrm{pc}^{-2}$). It follows that for a better insight in the distribution of the habitable stellar mass, at the disk areas with small stellar density, such as the outskirts, a higher mass resolution of the simulation is required.



\subsection{Limitations of the model}

While there are indications that late-epoch baryonic infall plays a significant role in realistic star-formation histories of any large spiral galaxy on both theoretical and observational grounds \citep[e.g.,][]{2001MNRAS.325..781F,2007ARA&A..45..221Bregman,2008JCAP...09..003P}, there are several justifications for not including this infall in the present simulation. First and foremost, together with results in \citet{2015arXiv151101786Forgan}, this is the first use of cosmological simulations of galaxy formation for astrobiological purposes, and it makes sense to make it as simple baseline case as possible. It is necessary to get at least some hold on the simplest, unperturbed case and its effective dynamics before proceeding to more complex perturbed histories. Second, we are dealing with a typical $\sim L_\ast$ spiral galaxy whose infall/merger history is unknown in advance, and the late infall is quite a complex process due to its intermittent and stochastic nature, immensely complicating the numerical task at hand. The fact that the infall history of the Milky Way is somewhat better understood prompts an astrobiologically important question how typical is Milky Way's star formation history in comparison to other similar systems \citep{2013MNRAS.435.2598K}. Third, suppression of star-formation below $z \simeq 0.5$ noted in IR surveys of Milky Way-like galaxies \citep{2015ApJ...803...26P} coupled with the quenched star formation in infalling dwarfs in the ViaLactea II simulations \citep{2012MNRAS.425..231R} taken together suggest that the impact of late infall on the astrobiologically interesting population of metal-rich disk stars is not large on the average. That said, an important future task for this kind of simulations, concurrent with increasing resolution, is to simulate such infall and its effects on radial mixing in particular.

\subsection{Comparison with other studies}

The present model results are in discrepancy with the studies of \citet{Gowanlock_et_al_2011AsBio}  and 
\citet{2015AsBio..15..683M}, who obtain much more centrally concentrated GHZ. While the 
exact reasons will require a separate study to ascertain, we find that one or more of the following 
played an important part in creating the discrepancy: {\bf (i)} the model of \citet{Gowanlock_et_al_2011AsBio}  
applies specifically to the Milky Way, and uses a semi-analytic stellar density and star-formation 
history models specifically geared toward the local observations. In contrast, the present study like 
the one of \citet{2015arXiv151101786Forgan} uses a full numerical treatment of a {\it Milky Way-like\/} giant spiral 
galaxy. {\bf (ii)} The metallicity gradient of \citet{2006MNRAS.366..899N} used by \citet{Gowanlock_et_al_2011AsBio} 
is shallower than the effective metallicity gradient obtained in the cosmological simulations, and is 
shallower than the best-fit global gradient obtained by \citet{2000A&A...358..869R} and \citet{2003NewA....8..737T}. 
The results of \citet{2000A&A...358..869R} were used in \citet{2012OLEB...42..347V}. {\bf (iii)} As 
Gowanlock et al. (2011) correctly note, their model does not take into account other mechanisms for 
decreasing habitability operational mostly in the inner parts of the disk/bulge of the Milky Way, 
notably dynamical disruption of stable planetary orbits, chemical evolution overshooting resulting in 
oceanic "super-Earths", ambient UV suppression of dust grains necessary to form habitable planets, 
rare, but extremely important $\gamma$-ray burst sterilizations, and even indirect effects such as 
the cosmic-ray flux-climate connection. Note that all these processes are likely to make the inner 
parts of the Galaxy less habitable than hitherto suspected.

\subsection{(A)typical Earth}
\begin{figure*}
\includegraphics[height=0.9\textheight]{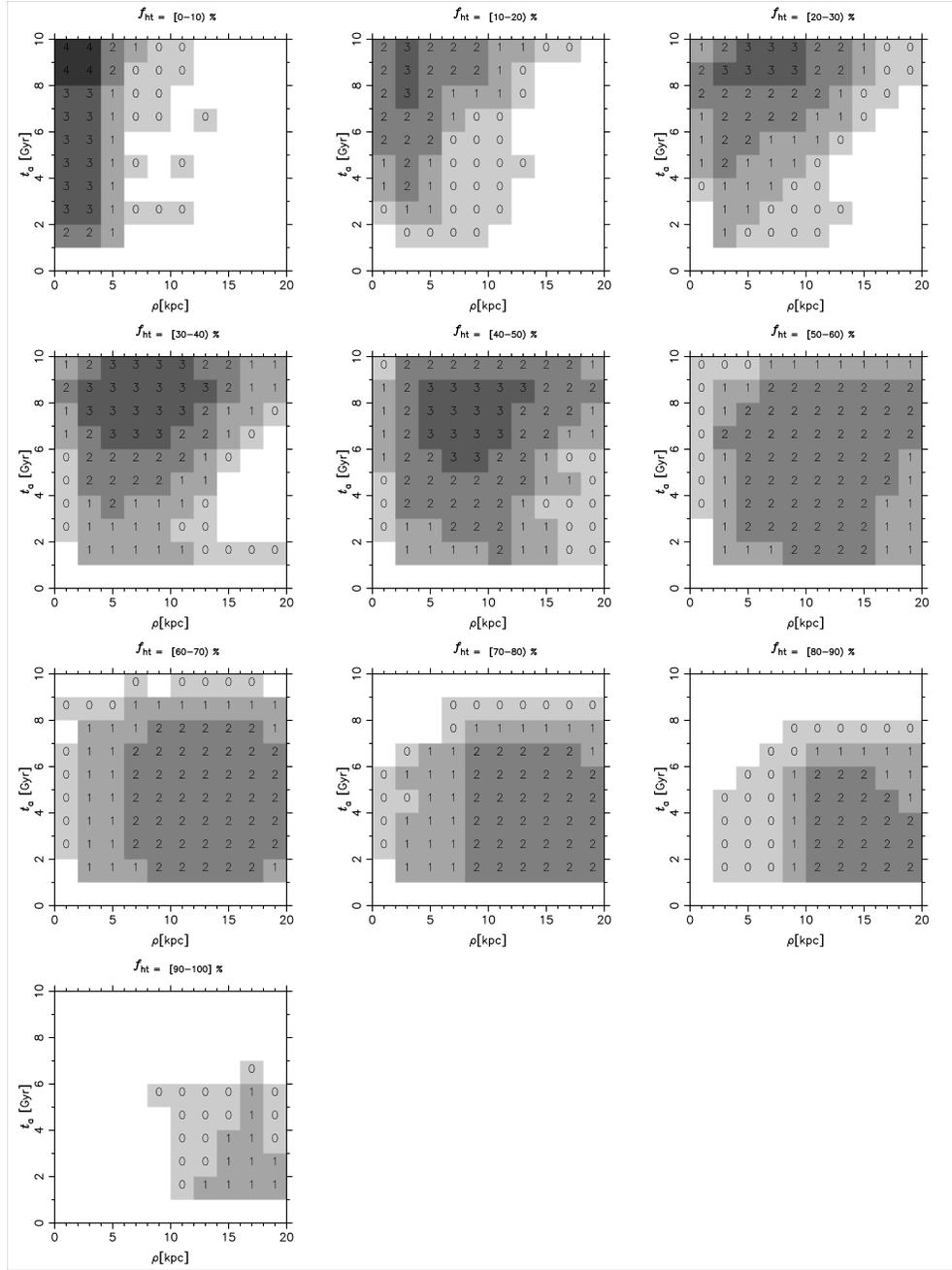}
\caption{CHPs distribution over $\rho$ and age for various
$f_\mathrm{ht}$ (intervals indicated in each panel). The numbers
are presented in order of magnitude form as both, shades of gray
and digits; Values $[0,10)$ with $0$, $[10,100)$ with $1$, $[100,1000)$ with $2$,
$[10^3,10^4)$ with $3$ and $[10^4,10^5)$ with $4$. Averaged over 5 runs.}
\label{rho_ht10}
\end{figure*}

For a more detailed analysis, the $t_\mathrm{h}$ distribution of
CHPs is examined. How typical is the Earth among habitable planets of
similar age and $f_\mathrm{ht}$?  In this modest attempt to answer
this formidable question we have constructed Figure
\ref{rho_ht10} and Table \ref{table1}.  In Table \ref{table1} we give the distribution
of CHPs number over $f_\mathrm{ht}$ and age. The peak values for
each given age are in bold font; We deliberately disregard the
values for $f_\mathrm{ht} < 10 \%$ since it is evident from
Figure \ref{rho_ht10} and their low $f_\mathrm{ht}$ that those
particles are likely members of bulge population with low
prospects for habitability. The CHPs that spent
almost all of their life time in the habitable friendly conditions
reside at $\sim16$ kpc from Galactic centre and are $\sim3$ Gyr
old (Figure \ref{rho_ht10}). A typical Solar system analog, $4-5$
Gyr old at $8-10$ kpc from Galactic centre, has the highest
probability for $f_\mathrm{ht} \sim 50\%$ (Figure \ref{rho_ht10}
and Table \ref{table1}). As the value for $f_\mathrm{ht}$ gets
higher the probability peak moves to higher $\rho$. At $f_\mathrm{ht} \ge 90 \%$ less than $1\%$ of the systems (that are $4-5$ Gyr old) can be found to reside at $8-10$ kpc from Galactic centre (Figure
\ref{rho_ht10} and Table \ref{table1}). For smaller values of $f_\mathrm{ht}$ this fraction is likely to be as high as $\sim 10 \%$. The highest value in bold typeface from Table
\ref{table1} is for particles with $30\% < f_\mathrm{ht} < 40\%$
old between $8$ and $9$ Gyr. From Figure \ref{rho_ht10} it is
evident that these particles are the most numerous at $\rho<10$ kpc distance
from Galactic centre. Summing up the values in Table \ref{table1} (without the values in $f_\mathrm{ht} < 10\%$ row), a typical CHP in the galactic disk has $\sim 35\%$ chances to be habitable ($f_\mathrm{ht} > 50\%$). With $f_\mathrm{ht} < 10\%$ row included, chances drop to $\sim 20\%$. This indicates the importance of SFR, stellar density and dynamics for the regulation of Galactic habitability.

\subsection{Wider GHZ?}

Figure \ref{hc_frac_rho} implies that areas in the $10-15$ kpc range have the highest probability of containing habitable systems and although the same order of magnitude habitability exists for most of the $\rho$ range, the area between $10-15$ kpc should be favoured in terms of habitability. This complies with the results on Galactic habitable zone in \citet{Pena_CabreraDurand_Manterola04}, where the zone of enhanced habitability probability is estimated to be at $\rho = (4, 17.5)$ kpc. Our results differ from $\rho = (7, 9)$ kpc conservative estimates in \citet{Lineweaver_et_al_2004Sci} probably caused by calibrating threshold habitability values of SFR and stellar density to Solar neighbourhood. Also our zone of enhanced habitability does not simply drift to the outskirts of the disk with time \citep[as also suggested in][]{Gonzalez_et_al_2001Icar} but rather have a rapid build-up at $\rho = (10,15)$ kpc after $5$ Gyr, caused by dynamical friction effects that scatter gas and stars away from the Galactic centre. In more realistic cases, these effects might be partially balanced by minor mergers, resupplying the central regions with gas and stars, which remains an interesting topic for future studies.

\begin{table*}
\caption{Number of CHPs distributed over $f_\mathrm{ht}$ and ages averaged over 5 runs. Values in bold type face are maxima for the given column, excluding values in (0-10) \% row.}
\begin{tabular}{cccccccccc}
 $ f_\mathrm{ht}[\%]$\textbackslash Gyr&       1-2&       2-3&       3-4&       4-5&       5-6&       6-7&       7-8&       8-9&      9-10\\ \hline
      0-10&       935&      1413&      1863&      1707&      2018&      3186&      7003&     14781&     55387\\
     10-20&        11&        44&       165&       383&       582&       823&      1258&      2142&      1355\\
     20-30&        23&        46&       120&       225&       453&       857&      1740&      4651&      4039\\
     30-40&        79&        90&       254&       598&      1342&      3028&\bf{  6642}&\bf{ 12694}&\bf{  6403}\\
     40-50&       303&       434&      1048&      1964&\bf{  3140}&\bf{  4291}&      5647&      5996&      1729\\
     50-60&       933&      1033&      1669&\bf{  2035}&      2580&      2875&      2369&      1494&       296\\
     60-70&      1603&\bf{  1840}&\bf{  1999}&      1986&      2115&      1755&      1003&       333&         7\\
     70-80&\bf{  1789}&      1735&      1582&      1468&      1190&       728&       181&        20&         0\\
     80-90&      1004&       901&       860&       648&       416&       101&        19&         0&         0\\
    90-100&       139&        55&        37&        36&        23&         2&         0&         0&         0\\
\end{tabular}
\label{table1}
\end{table*}

\subsection{Contact prospects}
These findings shed a new light on the importance and relevance of the GHZ concept on our astrobiological and SETI studies. In contrast to the simplistic approach of the early work on Galactic habitability, it seems now that whatever metric of habitability is adopted, the distribution over spatial locations and temporal history will be complex, with much hitherto unnoticed medium- and small-scale structure. Moreover, it could be expected that future higher-resolution studies will reveal still more structure on smaller scales. In contrast even to more sophisticated models based on chemical evolution \citep[e.g.][]{2014MNRAS.440.2588Spitoni_etal}, we perceive a higher fraction of the habitable set on higher galactocentric distances, beyond 10 kpc.

This might be in accordance with several alternative hypotheses suggested in SETI studies. In particular, regions further out in the Milky Way disk might be better targets for practical search programmes than previously suspected \citep[cf.][]{2006NewA...11..628CirkovicBradbury}. To what extent is the realistic GHZ porous enough to leave larger "persistent" bubbles capable of explaining Fermi's paradox \citep{1998JBIS...51..163Landis, 2001cond.mat.12137Kinouchi}, remains to be seen in future work of higher resolution.

At $0.9<f_\mathrm{ht}<1$ panel in Figure \ref{rho_ht10} it is likely for a planet with the highest chances of habitability  to be younger than Earth instead of being older \citep[as suggested in][]{Lineweaver2001Icar}. This is in accordance with our perceived absence of contact with extraterrestrials \citep[with the usual assumptions of realism, scientific naturalism, and gradualism, cf.][]{2009SerAJ.178....1Cirkovic}. Absence of the contact was also explained, due to influence of habitability agents, by arguments of astrobiological phase transition resulting from an expected  decrease in SFR \citep[see][]{2012OLEB...42..347V}. In our model SFR sharply decreases after $3$ Gyr which makes the results of this paper even more appealing in explaining the lack of perceiving aliens so far.

The simulations performed here are similar to ones recently
reported by \citet{2015arXiv151101786Forgan}, including very
similar mass resolution ($5 \times 10^4 \, M_\odot$ vs. $3.16
\times 10^4 \, M_\odot$) for stellar particles, and consequent
low spatial resolution. Many caveats systematically presented by
\citeauthor{2015arXiv151101786Forgan} apply here as well, notably the impossibility to
treat individual sterilization events due to local supernovae
with the state-of-the-art resolution. The present model
introduces the notion of continuous habitability, which changes
its results somewhat in comparison to those presented in
particular in Figure 7 of \citet{2015arXiv151101786Forgan}. On
the other hand, it is important to note a general similarity
between results on the Milky Way habitability obtained in two
independent studies, especially when it comes to the seemingly
counter-intuitive conclusion that outer edges of the Galactic
disk are more habitable than hitherto assumed. Clearly, future
improvements in both resolution and adding further dynamical
mechanisms (spiral-arm crossings, Galactic tides stemming from
vertical oscillations, $\gamma$-ray bursts, etc.) will further
constrain habitability and offer more meaningful target selection
for observational searches.

\section{Summary}
We present an isolated Milky Way simulation in GADGET2 N-body SPH
code to model the habitability of Galactic disk in order to
assess the question of perception of Earth-like extraterrestrial
habitats. The peak values for possibly habitable stellar mass
surface density move from $10$ to $15$ kpc galactocentric
distance in $10$ Gyr simulated time span. The stellar particles
that spent almost all of their life time in a habitable friendly
conditions reside at $\sim16$ kpc from Galactic centre and are 
 $\sim3$ Gyr old.  Stellar particles that are $4-5$ Gyr old with $f_\mathrm{ht} \ge 90 \%$, are also predominantly found in the outskirts of the Galactic disk. Less then $1 \%$ of these particles can be found at a typical Solar system galactocentric distance of $8-10$ kpc. Although we include the effects of
Galactic dynamics in our model our results appear strongly
dependent on distributions of SFR and stellar density resulting
from our simulation and imposed habitability criteria.
Examination of this dependence is a viable scope for future
studies and projects that will model changes in SFR and stellar
density parameters caused by interaction of a Galaxy with its
environment.

We argued in favour of a hypothesis that any practical search for
other Galactic complex biospheres/extraterrestrial civilizations
should be steered towards the outskirts of the Galactic disk and
stellar systems younger than ours. This is in accordance with our
present lack of SETI results. We may not be alone out there after
all, but we are older then they are. The room for future studies,
both observational and numerical is wide open (both literally and
metaphorically).

\section{Acknowledgements}

GMA and BV acknowledge financial support from ERASMUS-ASTROMUNDUS programme hosted by the University of Innsbruck where this paper was conceived in 2013. We thank an anonymous referee for useful comments that have greatly improved the quality of the paper. BV, MMC and MM acknowledge financial support from the Ministry of Education, Science and Technological Development of the Republic of Serbia through the project \#176021 "Visible and invisible matter in nearby galaxies: theory and observations".

\bibliographystyle{mn2e}
        \bibliography{mybib}

\end{document}